\documentclass{article}
\usepackage{fullpage}
\usepackage{graphics,psfrag,epsfig}
\usepackage{amsfonts,latexsym,eucal,amsmath,amsthm,amssymb}

\begin{document}

\newcommand{\rum}{\rule{0.5pt}{0pt}}
\newcommand{\rub}{\rule{1pt}{0pt}}
\newcommand{\rim}{\rule{0.3pt}{0pt}}
\newcommand{\numtimes}{\mbox{\raisebox{1.5pt}{${\scriptscriptstyle \times}$}}}
\newcommand{\optprog}[2]
{%
  \noindent\mbox{}\\[0cm]
  \noindent\fbox{%
  \begin{minipage}{0.955\linewidth}
    \mbox{}\\[-0.5cm]
    #1\\[#2]
  \end{minipage}
  }
  \noindent\mbox{}\\[-0.2cm]
}

\renewcommand{\refname}{References}

\twocolumn[%
\begin{center}
{\Large\bf An elegant argument that $\mathsf{P} \neq \mathsf{NP}$ \rule{0pt}{13pt}}\par
\bigskip
Craig Alan Feinstein \\ {\small\it 2712 Willow Glen Drive, Baltimore, Maryland
21209\rule{0pt}{13pt}}\\ \raisebox{-1pt}{\footnotesize E-mail: cafeinst@msn.com,
BS"D}\par
\bigskip\smallskip
{\small\parbox{11cm}{%
\bigskip \noindent \textbf{Abstract:} In this note, we present an elegant argument that $\mathsf{P} \neq \mathsf{NP}$ by demonstrating that the Meet-in-the-Middle algorithm must have the fastest running-time of all deterministic and exact algorithms which solve the SUBSET-SUM problem on a classical computer.

\bigskip \noindent \textbf{Disclaimer:} This article was authored
by Craig Alan Feinstein in his private capacity. No official support or endorsement by
the U.S. Government is intended or should be inferred.\rule[0pt]{0pt}{0pt}}}\bigskip

\bigskip \noindent ``This one's from \textit{The Book}!" - Paul
Erd\"{o}s (1913-1996)\bigskip\end{center}]{%

Consider the following problem: Let $\{s_1,\dots,s_n\}$ be a set of $n$ integers and $t$ be another integer. We want to determine whether there exists a subset of $\{s_1,\dots,s_n\}$ for which the sum of its elements equals $t$. We shall consider the sum of the elements of the empty set to be zero. This problem is called the SUBSET-SUM problem \cite{b:CLR90,b:MvOV96}. Let
$$
S_k^+=\bigg\{\sum_{i \in I^+} s_i \mid I^+ \subseteq \{1,\dots,k\}\bigg\}
$$
and
$$
S_k^-=\bigg\{\sum_{i \in I^-} s_i \mid I^- \subseteq \{k+1,\dots,n\}\bigg\},
$$
\noindent where $k \in \{1,\dots,n\}$. Notice that for any $k \in \{1,\dots,n\}$, the SUBSET-SUM problem is equivalent to the problem of determining whether set $S_k^+ + S_k^-$ intersects set $\{t\}$; therefore, for any $k \in \{1,\dots,n\}$, the SUBSET-SUM problem is equivalent to the problem of determining whether set $S_k^+$ intersects set $t-S_k^-$. Now consider the following algorithm for solving the SUBSET-SUM problem:

\bigskip\noindent \textbf{Meet-in-the-Middle Algorithm -}
Sort the sets $S_{\lfloor n/2 \rfloor}^+$ and $t-S_{\lfloor n/2 \rfloor}^-$ in ascending order. Compare the first elements in both of the lists. If they match, then output ``YES". If not, then compare the greater element with the next element in the other list. Continue this process until there is a match, in which case the computer outputs ``YES", or until one of the lists runs out of elements, in which case the computer outputs ``NO".

\bigskip This algorithm takes $\Theta(\sqrt{2^n})$ time, since it takes $\Theta(\sqrt{2^n})$ steps to sort sets $S_{\lfloor n/2 \rfloor}^+$ and $t-S_{\lfloor n/2 \rfloor}^-$ and $O(\sqrt{2^n})$ steps to compare elements from each of the two sets. It turns out that no deterministic and exact algorithm with a better worst-case running-time has ever been found since Horowitz and Sahni published this algorithm in 1974 \cite{b:HS74,b:Woe03}. We give a simple proof that it is impossible for such an algorithm to exist:

Let $k \in \{1,\dots,n\}$. Then the SUBSET-SUM problem is to determine whether there exist sets $I^+ \subseteq \{1,\dots,k\}$ and $I^- \subseteq \{k+1,\dots,n\}$ such that
$$
\sum_{i \in I^+} s_i = t - \sum_{i \in I^-} s_i.
$$
There is nothing that can be done to make this equation simpler. Then since there are $2^k$ possible expressions on the left-hand side of this equation and $2^{n-k}$ possible expressions on the right-hand side of this equation, we can find a lower-bound for the worst-case running-time of an algorithm that solves the SUBSET-SUM problem by minimizing $2^k+2^{n-k}$ subject to $k \in \{1,\dots,n\}$.

When we do this, we find that $2^k+2^{n-k}=2^{\lfloor n/2 \rfloor}+2^{n-\lfloor n/2 \rfloor}=\Theta(\sqrt{2^n})$ is the solution, so it is impossible to solve the SUBSET-SUM problem in $o(\sqrt{2^n})$ time; thus, because the Meet-in-the-Middle algorithm achieves a running-time of $\Theta(\sqrt{2^n})$, we can conclude that $\Theta(\sqrt{2^n})$ is a tight lower-bound for the worst-case running-time of any deterministic and exact algorithm which solves SUBSET-SUM. And this conclusion implies that $\mathsf{P} \neq \mathsf{NP}$ \cite{b:BC94,b:CLR90}.\qed

}


\smallskip
\begin{thebibliography}{99}\small

\bibitem{b:BC94} P.B. Bovet and P. Crescenzi, \textit{Introduction to the Theory of
    Complexity}, Prentice Hall, 1994.

\bibitem{b:CLR90} T.H. Cormen, C.E. Leiserson, and R.L. Rivest, \textit{Introduction to
    Algorithms}, McGraw-Hill, 1990.

\bibitem{b:HS74} E. Horowitz and S. Sahni, ``Computing Partitions with Applications to
    the Knapsack Problem", \textit{Journal of the ACM}, vol. 2l, no. 2, April 1974, pp
    277-292.

\bibitem{b:MvOV96} A. Menezes, P. van Oorschot, and S. Vanstone, \textit{Handbook of
    Applied Cryptography}, CRC Press, 1996.

\bibitem{b:Woe03} G.J. Woeginger, ``Exact Algorithms for NP-Hard Problems",
    \textit{Lecture Notes in Computer Science}, Springer-Verlag Heidelberg, Volume 2570,
    pp. 185-207, 2003.

\end{thebibliography}
\end{document}